\documentclass[prl,twocolumn,showpacs,floatfix]{revtex4}

\usepackage{epsf}
\usepackage{graphicx}
\usepackage{dcolumn}
\usepackage{bm}

\begin{document}
\bibliographystyle{apsrev}


 \flushbottom
 \renewcommand{\textfraction}{0}
 \renewcommand{\topfraction}{1}
 \renewcommand{\bottomfraction}{1}

\newcommand{\beq}{\begin{equation}}
\newcommand{\dd}{\partial}
\newcommand{\eeq}{\end{equation}}
\newcommand{\bea}{\begin{eqnarray}}
\newcommand{\eea}{\end{eqnarray}}



\preprint{UCLA/01/TEP/13} 

\title{Neutrino production in matter with time-dependent density or velocity
}

\author{Alexander Kusenko$^{1,2}$ and Marieke Postma$^1$
}
\affiliation{$^1$Department of Physics and Astronomy, UCLA, Los Angeles, CA
90095-1547 \\ $^2$RIKEN BNL Research Center, Brookhaven National
Laboratory, Upton, NY 11973 }

\date{July, 2001}

\begin{abstract}
\vspace{0.1cm}
We show that neutrinos can be produced through standard electroweak
interactions in matter with time-dependent density.  
  
\end{abstract}

\pacs{13.15.+g   \hspace{1.0cm} } 

\maketitle

\renewcommand{\thefootnote}{\arabic{footnote}}
\setcounter{footnote}{0}

Electroweak interactions couple neutrinos to matter fermions, which have
non-zero SU(2) charge.  Both ordinary matter and the nuclear matter of
neutron stars carry net SU(2) charge.  One can expect, therefore, that, if
the matter density changes with time, neutrinos can be pair-produced, in 
analogy with the photon production by a time-dependent density of electric
charge.  We will show that this is, indeed, the case.  

Neutrino pair-creation in dense matter is similar to the pair-creation of
fermions  by a variable electromagnetic field~\cite{book,grib,ct}
or by an oscillating scalar field~\cite{baa,gk}.  The mechanism we describe
is different from that discussed in~\cite{loeb}, where the matter density
gradients are essential.  In contrast, we will assume the density is
uniform and time-dependent.  The discussion below is
applicable to both ordinary matter and nuclear matter.  For definiteness, let
us consider the latter.   The weak interactions of
neutrons $n$ with neutrinos $\nu$ are described by the following potential:
\begin{equation}
{\cal L}_{\rm int} = (G_{F}/\sqrt{2}) \, [\bar{n} \gamma_\mu (1-\gamma_5)
n] \, [\bar{\nu} \gamma^\mu (1-\gamma_5) \nu],
\label{interaction} 
\end{equation}
where $G_F$ is the Fermi constant.  In nuclear matter, the current $j^\mu =
(G_{F}/\sqrt{2}) \langle \bar{n} \gamma^\mu (1-\gamma_5) n \rangle $, has a
non-zero expectation value.  For matter at rest, $j^\mu=(\rho,0,0,0)$; here
$\rho = (G_{F}/\sqrt{2})n_n$, and $n_n$ is the number density of neutrons.
For example, for matter densities $\rho_{M} \sim 10^{10}-10^{16} {\rm g}\,
{\rm cm}^{-3} $, inside a neutron star, the ``reduced density'' is $\rho \sim
10^{-3}-10^{3} \,{\rm eV}$.

If the matter density is time-dependent or if $\vec{j}$ changes direction, 
in the presence of interaction (\ref{interaction}), the neutrino field
satisfies the Dirac equation with a time-dependent potential: 
\begin{equation}
\left [ i \gamma^0  \dd_0 + i \gamma^k \dd_k +j_\mu(t)  \gamma^\mu
\frac{1-\gamma_5}{2} -m
\right ]  \psi =0.
\end{equation}

Let us consider the case of a uniform time-dependent density $\rho(t)$ 
moving with a four-velocity $v_\mu=\{v_0(t),0,0,v_3(t) \}$ along the third
axis. Then $j_\mu=\rho v_\mu$.  We will assume that both $\rho $ and
$v_\mu$ are constant at $t<0$, and become time-dependent for $t>0$.  We now
calculate the number of neutrinos produced by this variable nuclear
current.

Solutions of the Dirac equation can be written in terms of the eigenmodes: 
\beq
\psi = \sum_{s=\pm} \int \frac{d^3 k}{(2\pi)^3} 
\left [
b_{{\bf k},s} U_{{\bf k},s} (t) + d_{-{\bf k},s}^{\dag} V_{-{\bf k},s}(t) 
\right ] e^{i {\bf k}{\bf x} }, 
\eeq 
where the creation and annihilation operators satisfy the usual
anticommutation relations.  For the positive and negative energy modes we
use the {\em Ansatz}:
\beq
U_{{\bf k},s} (t)  = 
N_{{\bf k},s}^{(1)} \left [ i \gamma^\mu  \dd_\mu 
+\rho(t) v_\mu(t)  \gamma^\mu \frac{1-\gamma_5}{2} +m
\right ] F_{{\bf k},s}^{(1)}, 
\label{U}
\eeq
\beq
V_{-{\bf k},s} (t)  =  
N_{{\bf k},s}^{(-1)} \left [ i \gamma^\mu \dd_\mu
+\rho(t) v_\mu (t) \gamma^\mu \frac{1-\gamma_5}{2} +m
\right ] F_{{\bf k},s}^{(-1)}, 
\label{V}
\eeq
where 
\bea 
F_{{\bf k},s}^{(1)}& = & f^{(1)}_{k,s}(t)   \left ( 
\begin{array}{c}
\chi_s \\
0
\end{array}
\right )  \nonumber \\
F_{{\bf k},s}^{(-1)}& = & f^{(-1)}_{k,s}(t)  \left ( 
\begin{array}{c}
0 \\
\chi_s
\end{array}
\right ),  
\label{F}
\eea
and $\chi_\pm$ is an eigenspinor of $\vec{ k} \cdot \vec{ \sigma}$ with 
eigenvalue $\pm1$.  We will assume that vectors $\vec{v}$ and $\vec{k}$ are
aligned because it simplifies the calculations; we do not expect the
orthogonal momenta to change the overall picture of particle production. 

Substituting this {\em Ansatz} into the Dirac equation, we get the
equations for the mode functions $f^{(\lambda)}_{k,\pm}(t) $, $\lambda=\pm
1$.  We introduce
\beq 
\phi^{(\lambda)}_{k,\pm}(t) = \exp \left \{ -i/2 \int_0^t \rho_\pm (\zeta)
d\zeta \right \} f_{k,\pm}^{(\lambda)}(t),  
\label{phase} 
\eeq
which satisfies equation
\beq
\ddot{\phi}_{k,\pm}^{(\lambda)} + \left [
m^2 + \left (\frac{1}{2} \rho_\pm \pm k \right )^2  - i 
\frac{\lambda}{2} \dot{\rho}_\pm \right ] \phi_{k,\pm}^{(\lambda)} = 0, 
\label{mode}
\eeq 
where $\rho_\pm = \rho (v_0\pm v_3 )$ for $\vec{k}$ aligned with
the third axis.   Equation~(\ref{mode}) has two linearly independent
solutions ${\phi}_{k,+}^{(\lambda)}$ and
${\phi}_{k,-}^{(\lambda)}$ for $\lambda=1$.  Their complex conjugate
counterparts solve the same equation for $\lambda=-1$.  In other words,
\begin{equation}
\label{rel_phi}
\left (\phi_{k,\pm}^{(\lambda)} \right )^*= \phi_{k,\pm}^{(-\lambda)}. 
\end{equation}

In the presence of a matter background the charge conjugation 
symmetry C is spontaneously broken.  Equation~(\ref{rel_phi}) relates the
particles and antiparticles by a transformation that simultaneously changes
the sign of $\rho$ in equation~(\ref{phase}). 

\begin{figure}[t]
\centering
\hspace*{-5.5mm}
\leavevmode \epsfysize=5cm  \epsfbox{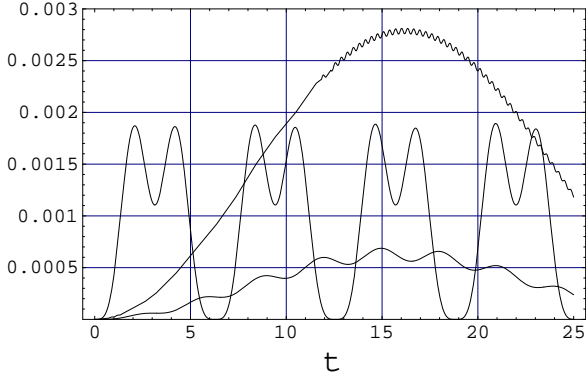}\\[3mm]
\caption[fig. 1]{\label{fig1} The occupation number of neutrino pairs, {$
{\mathcal N}_{{ k},+}^P$}, as a function of time, for $j^0 = 1+0.1
\sin(\omega t)$ and $\vec{j}=0$, all in units of $m$.  The upper curve
corresponds to $(\omega,k) = (10,4.3)$ (close to resonance), the middle
curve to $(1,-0.5)$, and the lower curve to $(0.1,-0.5)$.}
\end{figure}

The initial state is specified at $t \leq 0$, when the density is
time-independent: $\dot{\rho}_\pm(t\leq 0) = 0$.  In a constant density
background, the mode functions are plane waves, and we can choose the
following initial conditions: 
\beq
	f_{k,\pm}^{(\lambda)}(0) = 1, \qquad
	\dot{f}_{k,\pm}^{(\lambda)}(0) = - i E_{k,\pm}^{(\lambda)}(0).
\label{ini}	
\eeq
Here 
\beq
	 E_{k,\pm}^{(\lambda)}(0) = - \frac{1}{2} \rho_{\pm}(0) + \lambda 
	\sqrt{m^2 + \left (\frac{1}{2} \rho_{\pm}(0)\pm k \right )^2}.
\label{E0}
\eeq
are the eigenvalues of the Hamiltonian in matter with a constant density.
In the limit $\rho \to 0$, the spinors $U$ and $V$ become the usual
free field positive and negative frequency solutions.  We normalize
the spinors
\beq
\label{normalization}
	U_{{k},r}^{\dag}(0) U_{{k},s}(0) =  
	V_{-{k},r}^{\dag}(0) V_{-{k},s}(0) = \delta_{rs}.
\eeq
Due to initial conditions, the orthogonality relation
\beq
	 U_{{k},r}^{\dag}(0) V_{-{k},s}(0) = 0
\eeq
is satisfied automatically.  Since the time evolution is induced by a 
hermitian Hamiltonian, the normalization and
orthogonality properties of the spinors are preserved.  Condition
(\ref{normalization}) fixes the normalization constants:
\beq
\left ( N_{{k},\pm}^{(\lambda)} \right )^{-2} 
= \left( E^{(-1)}_{k,\pm}(0)   
\mp k \right )^2 + m^2. 
\label{N}
\eeq
To simplify notation we will drop the $\lambda$ label from the
normalization constants, which are equal for $\lambda=\pm1$.

The particle number $n_{k,s}(t)=\langle b^{\dag}_{k,s}(t) b_{k,s}(t)
\rangle $ is defined in terms of creation and annihilation operators.
However, when $b^{\dag}_{k,s}$ and $b_{k,s}$ are time-dependent, the
Hamiltonian does not, in general, remain diagonal at all times.  This is a
sign of particle production.  To determine the number of particles, one has
to diagonalize the Hamiltonian by a Bogoliubov transformation
or, equivalently, project the states at time $t$ from the basis of
mode functions $f(t)$ to that of plain waves.  A detailed discussion
and various applications of this standard technique can be found in
Refs.~\cite{book,grib,baa,gk}.

\begin{figure}[t]
\centering
\hspace*{-5.5mm}
\leavevmode\epsfysize=5cm \epsfbox{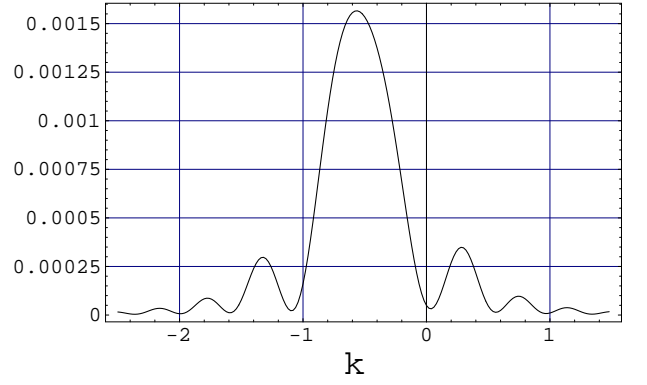}\\[3mm]
\caption[fig. 2]{\label{fig2} The number density of produced neutrino
pairs, {${\mathcal N}_{{k},+}^P$}, as a function of $k$ at $t = 10$.
The parameters chosen for this plot are $j^0 = 1 + 0.1 \sin(\omega t)$,
$\vec{j}=0$, and $\omega = 1$, all in units of $m$.  }
\end{figure}

We define the particle number in the basis of plain wave states in the
background of constant density $\rho_\pm=\rho_\pm(0)$.  The positive
and negative energy modes are given by eqs.~(\ref{U},\ref{V},\ref{F})
with energies (\ref{E0}) and normalization (\ref{N}).  In this basis
the time-dependent particle number is 
\beq
	{\mathcal N}^{P}_{k,s}(t) =|D^{P}_{k,s}|^2, 
\label{bog}
\eeq
where $D^{P}_{k,s}$ is the corresponding Bogoliubov coefficient:  
\bea
\label{Dpart}
	D^{P}_{k,\pm} = && {U_{{\bf k},\pm}}^{\dag}(0) V_{-{\bf k},\pm}(t)
		\\
	= && {m}{N_{{\bf k},\pm}^{2}}  
	\left(i \dot{f}^{(-1)}_{k,s}(t) - f^{(-1)}_{k,s}(t)
	E^{(-1)}_{k,\pm}(0)  \right ) .  \nonumber
\eea
Therefore, the particle number is 
\bea
{\mathcal N}^{P}_{k,\pm}(t) = && \frac{m^2}{\left[
(E_{k,\pm}^{(-1)}(0) \mp k)^2 + m^2 \right]^2}  \times \nonumber  \\
&& \left \{ |\dot{f}^{(-1)}_{k,\pm}(t)|^2 + 
| E^{(-1)}_{k,\pm}(0)
f^{(-1)}_{k,\pm}(t)|^2 - \right . \nonumber \\
&& \left . 2 E^{(-1)}_{k,\pm}(0) {\rm Im}
(f^{(-1)}_{k,\pm}(t) \dot{f}^{(-1)*}_{k,\pm}(t) )
\right \}.
\label{N(t)}
\eea

\begin{figure}[t]
\centering
\hspace*{-5.5mm}
\leavevmode\epsfysize=5cm \epsfbox{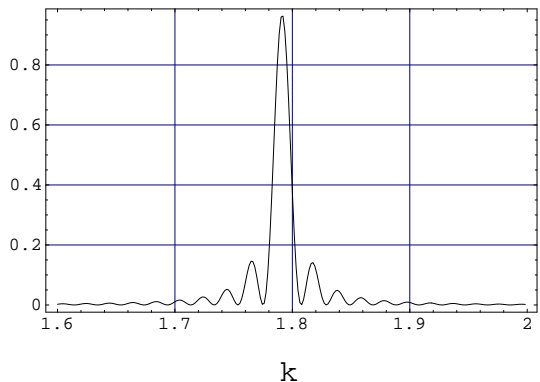}\\[3mm]
\caption[fig. 3]{\label{fig3} The occupation number for neutrino
pairs, {${\mathcal N}_{{k},+}^P$}, as a function of $k$ at $t = 175$.
The parameters are $j^\mu = 1+0.1 \sin(\omega t)$, $\vec{j}=0$ and
$\omega = 5$, all in units of $m$. The modes around $k \approx 1.79$
are in resonance.}
\end{figure}

In all formulas the neutrino momentum only enters in the combination
$k_{{\rm eff},\pm} (t) = k \pm \rho_\pm(t)/2$.  The momentum is shifted
due to the effective potential generated by the matter background.

The number of particles at $t=0$ is equal to zero due to the initial
conditions~(\ref{ini}) and (\ref{E0}). Furthermore,
equation~(\ref{rel_phi}) implies that at all times the number of
particles ${\mathcal N}^{P}_{k,s}(t) $ is equal the number of
antiparticles, ${\mathcal N}^{AP}_{-k}(t) =|D^{AP}_{-k}|^2 =
|{V_{-{k},\pm}}^{\dag}(0) U_{{k},\pm}(t) |^2$, as it should be.  The
total number density of neutrinos pairs is
\begin{equation}
	n_\nu (t) = \sum_{s=\pm} \int \frac{{\rm d}^3 k} {(2\pi)^3} 
	{\mathcal N}^{P}_{k,s}(t). 
\label{n_density}
\end{equation}
In the limit of an infinite volume the net amount of neutrinos
produced is $n_\nu(t_{\rm end}) - n_\nu(t_0)$, which goes to zero if
the matter density returns to its original state.  However, in reality
the matter with time-dependent density occupies a finite volume, so
that neutrinos can also diffuse away from the production region and
break the coherence that was implicitly assumed in derivation of
eq.~(\ref{N(t)}).  If the diffusion is fast, the density of neutrinos
is smaller than the maximal equilibrium value~(\ref{N(t)}) and is
nearly constant.  In this regime, one can define a particle production
rate $\dot{n}_\nu(t)$~\cite{book}.

Particle production vanishes in the limit of a massless neutrino, see
eq.~(\ref{N(t)}). This is easy to understand in the current formalism.
In the massless limit the right and left handed neutrino sectors
decouple, and if the Hamiltonian is diagonal at the initial time it
remains so at all times.  The Boguliubov transformation is trivial, and,
hence, no particle production occurs.  This is a consequence of the
perturbation $\delta \rho(t)$ being spatially homogeneous.  
The neutrinos are produced in pairs, and the two neutrinos have opposite
momenta in the center of mass frame. 
Thus, the matrix elements of the form $\langle
\bar{\nu}_{\!-\vec{k}} \Gamma^\mu \nu_{\vec{k}} \rangle$, where $
\Gamma^\mu = \gamma^\mu, \gamma^\mu \gamma^5$, vanish for massless
neutrinos.  Spatial perturbations $\delta \rho (\vec{x},t)$ lead
to a particle production only if they are time-dependent in all Lorentz
frames.  For example, for a plane wave perturbation $\delta \rho
\propto \cos(\vec{k} \cdot \vec{x} -\omega t)$ one can always make a
Lorentz transformation such that $\delta \rho' \propto \cos(\vec{k'}
\cdot \vec{x})$, and particle number remains constant in time.  Here
the phonons are virtual, and pair production is only possible if
additional energy is pumped in the system, as is done in the
production mechanism described in \cite{loeb}.
S
Using the mode equations (\ref{mode}), we can compute the first and
second derivatives of ${\mathcal N}^{P}_{k,\pm}$ at $t=0^+ $.  Let us
consider time evolution of a two-component vector ${\mathcal F} = \{
f^{(-1)}_{k,\pm}(t), \dot{f}^{(-1)}_{k,\pm} \}$.  One can rewrite the
mode equation as
\beq
\frac{d}{dt} {\mathcal F} = S {\mathcal F}, 
\eeq
where $S$ is a 2$\times$2 matrix: 
\beq
S= \left (
\matrix{
0 & 1 \cr
-(k^2 \pm k \rho_\pm + m^2) & i \rho_\pm(t) 
}
\right ).
\eeq 

The expression (\ref{N(t)}) can also be rewritten as 
\beq
{\mathcal N}^{P}_{k,\pm}(t) = m^2 N_{{k},\pm}^4 {\mathcal F}^{\dag} (t) T {\mathcal F} (t),
\label{N(t)matr}
\eeq
where 
\beq
T = \left (
\matrix{
|E^{(-1)}_{k,\pm}(0)|^2 & i E^{(-1)}_{k,\pm}(0) \cr
-i  E^{(-1)}_{k,\pm}(0) & 1
}
\right ).  
\eeq 
Now we differentiate eq.~(\ref{N(t)matr}) and use the
identity $T {\mathcal F}(0)=0$: 
\bea 
\dot{\mathcal N}^{P}(0) & = & m^2 N_{{k},\pm}^4 
{\mathcal F}^{\dag}(0) (S^{\dag} T + T S) {\mathcal F}(0) = 0;  \nonumber \\
\ddot{\mathcal N}^{P}(0) &= & m^2 N_{{k},\pm}^4 
{\mathcal F}^{\dag}(0) [ S^{\dag} (S^{\dag} T + T S) +  \nonumber \\
& & (S^{\dag} T + T S) S + 
( \dot{S}^{\dag} T + T \dot{S}) ] {\mathcal F}(0) \nonumber \\ & = & 
2 m^2 N_{{k},\pm}^4 \ {\mathcal F}^{\dag}(0) S^{\dag} T S {\mathcal F}(0).  
\eea
Finally,
\beq
\dot{\mathcal N}^{P}_{k,\pm}(0) =  0;   \ \ 
\ddot{\mathcal N}^{P}_{k,\pm}(0)  =  \frac{1}{2} m^2 N_{{k},\pm}^4 
\dot{\rho}_\pm^2(0).   
\label{dd} 
\eeq

Let us now consider a spatially uniform matter density that is
changing periodically with time:
\beq
j_\mu(t) = \rho(t) v_\mu, \ v_\mu = {\rm const}. 
\eeq
Figures 1-3 show the numerical solutions for the case $j^0/m = 1+0.1
\sin(\omega t)$, $\vec{j}=0$ and various values of $(\omega, k)$. The
occupation number density (\ref{N(t)}) is a periodic function of time,
as shown in Fig. 1. This means that neutrinos are created and absorbed
through non-perturbative interactions with matter ({\em
cf}. Ref.~\cite{book}).  At all times ${\mathcal N}^{P}_{k,\pm}(t)$
remains smaller than 1, in accordance with the Pauli exclusion
principle.  The neutrino pair-production occurs in a spectral band
centered around $k_{\rm eff}(t_0) = k \pm\rho_\pm(t_0)/2=0$, as shown
in Fig.~\ref{fig2}.  The value $k_{\rm eff}(t_0) =0 $ minimizes the
real part of the frequency in eq.~(\ref{mode}). In the adiabatic
regime, where $\dot{\rho} \ll k_{\rm eff}^2 + m^2$, the width of the
resonance peak is $\Gamma \sim m$. This clearly shows the
non-perturbative nature of neutrino pair production in a coherent
background.  A perturbative calculation involving multiple phonons,
each decaying into a neutrino pair, would predict a peak in the
spectrum around $k_{\rm eff} \sim \omega/2$ with width $\Gamma \sim
G_{\rm F}^2 \omega^3$ for $\omega > m$.  

For oscillations with $\rho(t) = \rho_0 + \delta \rho \sin (\omega t)$,
$\delta\rho \ll \rho_0$, equation (\ref{mode}) is a (complex) Mathieu
equation with purely imaginary parameter $q=i \delta\rho/\omega$ and the
$A$-parameter that is also time dependent: $A(k,t)=4(m^2+(\rho_\pm(t)/2 \pm
k)^2)/\omega^2$.  Strictly speaking, this time dependence makes equation
(\ref{mode}) different from the usual Mathieu equation, but the general
behavior of its solutions is the same for $\delta \rho \ll \rho$.  
Since the $q$-parameter is purely
imaginary, all solutions are semi-periodic.  There are several resonances,
the first (and the widest) of which occurs when $A=1$, that is when the
frequency and the momentum are related by $4 \omega^2 \approx m^2
+ (k \pm \rho_0/2)^2 $.
Fig. 3 shows this resonance for $\omega^{\rm res} =5 m$, at time $t = 175
m^{-1}$.  The occupation number ${\mathcal N}^{P}_{k,\pm}$
grows for those modes that are in resonance.

There are additional resonant bands for higher values of $A$, that is for
smaller oscillation frequencies.  However, since the ratio $q/A \propto
\omega$ decreases with $\omega$, the higher resonances have diminishing
width.  Dense stars and other astrophysical objects are generally
characterized by small frequencies, so that neutrino production occurs
off-resonance. 

The parameters in Fig. 1-3 were chosen so as to illustrate the 
general behavior of the solutions.  We now examine the realistic values of
these parameters in the neutron stars.  

Unless the lightest neutrino is much lighter than the $0.1$~eV scale
inferred from the mass-squared
difference measured by Super-Kamiokande, the basic
oscillation frequencies in neutron stars have $\omega \ll m$, very far from
the lowest - and widest - resonance.  Of course, higher modes of
oscillation may be excited, but they are usually associated with
significant dissipation.  Let us consider the efficiency of neutrino
production for $\rho(t) = \rho_0 + \delta \rho \sin (\omega t)$ with a
small driving frequency $\omega \ll m$.  Based on expansions of the
Bogoliubov coefficients in eqs. (\ref{Dpart},\ref{N(t)}), we estimate that
the neutrino energy production per unit volume is $\dot{\epsilon} \sim m^2
\max [m^2,(\delta \rho)^2] \omega$.  For a realistic case $m\sim \delta
\rho \sim 0.1$eV, one can estimate the neutrino production rate in a
neutron star with radius $R\approx 10$~km for the basic oscillation
frequency $\omega \sim 2 \pi c/R$:
\begin{equation}
\dot{E} \sim 10^{22} \left ( \frac{m}{0.1 {\rm eV}} \right
)^2 \left ( \frac{\max [m,\delta \rho]}{0.1 {\rm eV}} \right
)^2 
{\rm erg \ s}^{-1}
\label{Eout}
\end{equation}
This corresponds to emission of $10^{34}$ low-energy (sub-eV) neutrinos per
second.  For smaller amplitudes or larger frequencies, the production rates
decrease or increase, respectively.  

This effect is, clearly, negligible when compared to the energy scale of a
supernova explosion.  However, it is conceivable that future observations of
neutron stars may reveal subtle changes of the rotation speeds of cold
oscillating neutron stars that can be compared with our predictions. 
Neutrino production through the same mechanism can also occur in 
a neutron star binary system.  In the latter case pair-production
occurs because the centripetal acceleration $a=\dot{\vec{v}} \neq 0$,
and the current $j_\mu(t)$ has a non-zero derivative.  

In summary, we have described pair creation of neutrinos in a
background with time-dependent matter density.  

We thank N.~Graham, Y.~Levin, and R.~Peccei for very helpful discussions.
This work was supported in part by the US Department of Energy grant
DE-FG03-91ER40662, Task C, as well as by a Faculty Grant from UCLA Council
on Research.


\end{document}